\documentclass[
 reprint,superscriptaddress,
 amsmath,amssymb,prl,
 aps]{revtex4-1}

\usepackage{graphicx}
\usepackage{dcolumn}
\usepackage{bm}
\usepackage[ruled]{algorithm2e}
\usepackage{multirow,booktabs}
\usepackage{epstopdf}
\usepackage{svg}
\usepackage{paralist,siunitx,soul}
\usepackage[version=4]{mhchem}
\usepackage{hyperref}
\usepackage{url}

\hypersetup{colorlinks,urlcolor=blue,citecolor=blue,linkcolor=blue}
\DeclareUrlCommand\url{\color{blue}}

\begin{document}

\title{Effect of Concentration Fluctuations on Material Properties of Disordered Alloys}

\author{Han-Pu Liang}
\affiliation{Eastern Institute of Technology, Ningbo 315200, China}

\author{Chuan-Nan Li}
\affiliation{Materials Department, University of California, Santa Barbara, California 93106-5050, USA}

\author{Xin-Ru Tang}
\affiliation{Beijing Computational Science Research Center, Beijing 100193, China}

\author{Xun Xu}
\affiliation{Beijing Computational Science Research Center, Beijing 100193, China}

\author{Chen Qiu}
\affiliation{Eastern Institute of Technology, Ningbo 315200, China}

\author{Qiu-Shi Huang}
\affiliation{Eastern Institute of Technology, Ningbo 315200, China}

\author{Su-Huai Wei}
\email{suhuaiwei@eitech.edu.cn}
\affiliation{Eastern Institute of Technology, Ningbo 315200, China}


\begin{abstract}
	 Alloying compound $AX$ with another compound $BX$ is widely used to tune material properties. For disordered alloys, due to the lack of periodicity, it has been challenging to calculate and study their material properties. Special quasi-random structure (SQS) method has been developed and widely used to treat this issue by matching averaged atomic correlation functions to those of ideal random alloys, enabling accurate predictions of macroscopic material properties such as total energy and volume. However, in $A_xB_{1-x}$ alloys, statistically allowed local concentration fluctuations can give rise to defect-like minority configurations, such as bulk-like $AX$ or $BX$ regions in the extreme, which could strongly affect calculation of some of the material properties such as semiconductor band gap, if it is not defined properly, leading to significant discrepancies between theory and experiment. In this work, taking the bandgap as an example, we demonstrate that the calculated alloy bandgap can be significantly underestimated in standard SQS calculations when the SQS cell size is increased to improve the structural model and the band gap is defined conventionally as the energy difference between the lowest unoccupied state and the highest occupied state, because the rare event motifs can lead to wavefunction localization and become the dominant factor in determining the "bandgap", contrary to experiment. To be consistent with experiment, we show that the bandgap of the alloy should be extracted from the majority configurations using a density-of-states fitting (DOSF) method. This DOSF approach resolves the long-standing issue of calculating electronic structure of disordered semiconductor alloys. Similar approaches should also be developed to treat material properties that depends on localized alloy wavefunctions.
\end{abstract}

\maketitle

Semiconductor alloys $A_{1-x}B_{x}X$, formed by mixing compounds $AX$ and $BX$, are technologically significant because their structural, electronic, and optical properties can be tuned by adjusting the alloy composition $x$ or atomic configurations \cite{2008-PRL-GaInP-alloy,2014-PRB-perovskite,2022-PRL-Nitride-alloy,2022-NP-AgBiS2,2023-NCS-defect-alloy-review,2024-Nature-SQS,2024-PRL-disorder-Ga2O3}.
However, accurate description of the properties of disordered alloys is computationally challenging, as the disordered alloys lack perfect periodicity, making conventional band structure calculation approaches difficult to apply. The special quasi-random structure (SQS) method \cite{1990-PRB-SQS,1990-PRL-SQS,atat-sqs,1992-PRL-LRO} addresses this challenge by constructing finite supercells that mimic the ensemble-averaged atomic correlation functions of an ideal random alloy. This approach has been successfully applied to predict the random alloy properties and extended to study partially disordered alloys, such as structural properties of Zn$_x$Mg$_{1-x}$S$_y$Se$_{1-y}$ alloys, thermodynamical properties of high-entropy alloys, and magnetic properties in dilute magnetic semiconductors \cite{1998-PRL-LiCoO2,1998-PRL-ZnMgSSe,2004-PRB-dilute,2019-PRM-CuGaSe2-disorder,2024-JACS-ABX2}, where these properties are largely governed by global atomic environments. 

However, in $A_{1-x}B_{x}X$ alloys, while the global average concentration is $x$, local concentration fluctuations naturally occur, with statistically-allowed local region concentration varying between 0 and 1. In small SQS supercells, extreme configurations are unlikely to appear, but as the cell size increases, local regions with extreme concentrations can emerge, effectively forming defect-like configurations due to their low probability. For quantities like energy and volume, these local fluctuations have negligible effect on the overall statistical properties. For non-statistical quantities such as the energy level of the highest occupied state (HOS) and lowest unoccupied state (LUS) which are sensitive to the local configurations, the minority configurations can lead to wavefunction localization and become the dominant factor in determining the bandgap if it is defined as the energy difference between the LUS and HOS, i.e., $E_g^{\text{LUS-HOS}}$, leading to reduced or even vanishing bandgap \cite{2011-NM-disorderalloy,2020-ACS-bandtail,2020-PRB-bandtail,2022-ACSO-alloy-review}, in contrast with experimental observations. This is because the experimental bandgap measurements reflect the electronic states associated with the majority atomic environments \cite{1968-MRB-Tauc}, not the band tails induced by rare atomic motifs.

\begin{figure}[!b]
	\includegraphics[width=8.5cm]{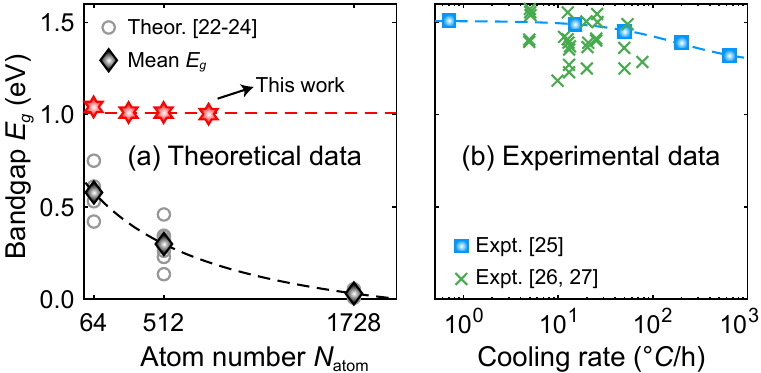}
	\caption{\label{fig:fig1} (a)Theoretical bandgap $E_g^{\text{LUS-HOS}}$ as a function of atom number $N_{\text{atom}}$ in a SQS cell and (b) experimental measured bandgap as a function of cooling rate in disordered \ce{Zn_{0.5}Sn_{0.5}P} alloys. The theoretical and experimental bandgap values are from Ref. \cite{2014-PRB-ZnSnP2,2012-APL-ZnSnP2-disorder,2015-PRB-unconvergence,2017-JPCC-ZnSnP2-expt-LRO,1987-JMR-ZnSnP2-expt,2015-PSSC-ZnSnP2}. Red hexagram marker denotes the bandgap from this work.}
\end{figure}

Taking random \ce{Zn_{0.5}Sn_{0.5}P} alloy as an example, conventional calculations shows poor convergence and even closure of the band gap as the supercell size increases \cite{2014-PRB-ZnSnP2,2012-APL-ZnSnP2-disorder,2015-PRB-unconvergence}. 
Experimentally, different cooling rates are used in growth to produce partially disordered \ce{Zn_{0.5}Sn_{0.5}P} alloys with different ordering parameters \cite{2017-JPCC-ZnSnP2-expt-LRO,1987-JMR-ZnSnP2-expt,2015-PSSC-ZnSnP2}. The measured bandgaps are within a range from 1.2 to 1.5 eV, as shown in Fig. \ref{fig:fig1}(b). Although the alloys are typically not in random state, it is clear that the measured values deviate significantly from the theoretical calculated values. 
Therefore, it is essential to clarify the role of concentration fluctuations in disordered alloy simulations and to develop realistic approaches to calculate alloy bandgaps. 

In this Letter, we highlight the occurrence and importance of the defect-like minority configurations arising from concentration fluctuations and show that, despite their low probability, they can have a profound impact on electronic properties, such as the HOS and LUS as well as the so defined bandgap $E_g^{\text{LUS-HOS}}$. From the perspective of majority configurations and based on a well-established shape of the density of states,  we propose a density-of-states fitting (DOSF) approach to define the bandgap of the disordered alloy, aligning theoretical definitions with experimental observations as well as ordered alloys. Applying this method to random \ce{Zn_{0.5}Sn_{0.5}P} alloy, we obtain a stable and converged bandgap of approximately 1.0 eV, independent of supercell size, significantly improving upon previous theoretical results. Furthermore, we explore partially disordered alloys with varying degrees of long-range order, achieving excellent agreement with experimental values. This work resolves a long-standing challenge in accurately modeling concentration-fluctuation-dependent properties of disordered alloys and offers a pathway for designing disordered semiconductors for advanced technological applications.

In disordered alloys, the random distribution of atoms leads to diverse local structural motifs. For example, in a random $A_{1-x}B_{x}X$ alloy, the anion-centered nearest-neighbor tetrahedra exhibit five motifs, from $A_4B_0$ to $A_0B_4$, with respective occurrence ratios being 1:4:6:4:1, as shown in Tab. \ref{tab:tab1}.
For second-nearest-neighbor polyhedra, there are 13 motifs, among which $A_{12}B_0$ and $A_0B_{12}$ occur with a low probability of 1/4096. Thus, the probability to have both Zn or both Sn atoms at the nearest and next nearest neighbor cluster is very small, at 1/65536. 
When more distant polyhedra are considered, the occurrence probability approaches zero. Such rare motifs inevitably appear because of statistical fluctuations. Due to their low probability, these minority configurations become defect-like and can introduce localized states near the band edges, significantly reducing the theoretical "bandgap" if it is not defined properly.

\begin{figure}[!b]
	\includegraphics[width=8.5cm]{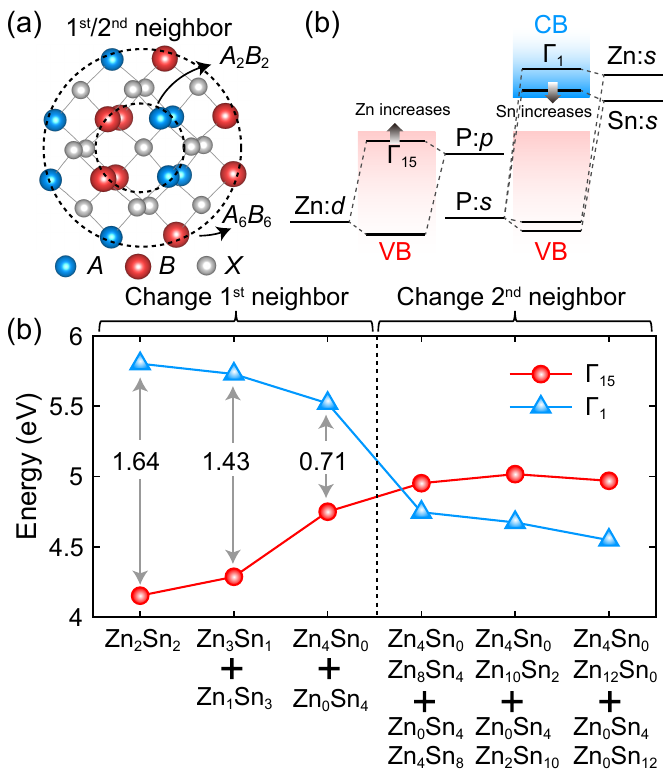}
	\caption{\label{fig:fig2} (a) Schematic of the nearest-neighbor tetrahedron $A_2B_{2}$ and second-nearest-neighbor polyhedron $A_6B_{6}$. (b) Orbital coupling diagram of the \ce{ZnSnP2} compound. (c) Band alignment of $\Gamma_{15}$ and $\Gamma_1$ showing the effect of varying nearest-neighbor tetrahedra from \ce{Zn2Sn2} to \ce{Zn4Sn0{/}Zn0Sn4}, and second-nearest-neighbor polyhedra from \ce{Zn6Sn6} to \ce{Zn12Sn0{/}Zn0Sn12}, with the nearest tetrahedra fixed at \ce{Zn4Sn0{/}Zn0Sn4}.}
\end{figure}

\begin{table*}[!htb]
	\caption{Relative distributions of first- and second-nearest-neighbor anion-centered polyhedra in random 64-atom and 512-atom $A_{0.5}B_{0.5}X$ supercells, as well as in partial disordered supercells with long-range order $\eta=0.25, 0.5,$ and $0.75$. For comparison, the relative distributions of ideal random and partially ordered alloys are also given in the table.}\label{tab:tab1}
	\begin{ruledtabular}
		\renewcommand{\arraystretch}{1.4}
		\begin{tabular}{lccccccccccc}
			& & \multicolumn{3}{c}{Random} & \multicolumn{2}{c}{$\eta=0.25$}& \multicolumn{2}{c}{$\eta=0.5$}& \multicolumn{2}{c}{$\eta=0.75$} & Order \\
			\cline{3-5} \cline{6-7}\cline{8-9} \cline{10-11}\cline{12-12}
			& Polyhedra & Ideal & 64 atoms  & 512 atoms & Ideal & Calc. &  Ideal & Calc. & Ideal & Calc. & Ideal \\
			\hline
			\multirow{5}{*}{1$^{\text{st}}$ nearest} & $A_0B_4$ & 1 & 1 & 1 & 0.88 & 0.88 & 0.56 & 0.56 & 0.19 & 0.19 & 0 \\
			& $A_1B_3$ & 4 & 4 & 4 & 3.98 & 4.00 & 3.75 & 3.75 & 2.73 & 2.81 &  0 \\
			& $A_2B_2$ & 6 & 6 & 6 & 6.27 & 6.25 & 7.38 & 7.38 & 10.15 & 10.00 & 16\\
			& $A_3B_1$ & 4 & 4  & 4 & 3.98 & 4.00 & 3.75 & 3.75 & 2.73 & 2.81 & 0\\
			& $A_4B_0$ & 1 & 1  & 1 & 0.88 & 0.88 & 0.56 & 0.56 & 0.19 & 0.19 & 0\\
			\hline
			\multirow{13}{*}{2$^{\text{nd}}$ nearest} & $A_0B_{12}$ & 0.06 & 0 & 0.00 & 0.04 & 0.00 & 0.01 & 0.00 & 0.00 & 0.00 & 0\\
			& $A_1B_{11}$ & 0.75 & 0 & 0.50 & 0.58 & 0.00 & 0.22 & 0.50 & 0.02 & 0.00 & 0\\
			& $A_2B_{10}$ & 4.13 & 0 & 3.50 & 3.52 & 3.50 & 1.92 & 2.00 & 0.34 & 0.50 & 0\\
			& $A_3B_9$ & 13.75 & 16  & 11.50 & 12.76 & 14.00 & 9.35 & 8.00 & 3.29 & 3.00 & 0\\
			& $A_4B_8$ & 30.94 & 48  & 32.50 & 30.52 & 31.00 & 28.18 & 22.00 & 18.46 & 21.00 & 0\\
			& $A_5B_7$ & 49.50 & 48  & 50.00 & 50.65 & 50.00 & 54.41 & 61.50 & 58.66 & 57.00 & 0\\
			& $A_6B_6$ & 57.75 & 32  & 60.00 & 59.83 & 59.00 & 67.80 & 68.00 & 94.46 & 93.00 & 256\\
			& $A_7B_5$ & 49.50 & 48  & 50.00 & 50.65 & 50.00 & 54.41 & 61.50 & 58.66 & 57.00 & 0\\
			& $A_8B_4$ & 30.94 & 48  & 32.50 & 30.52 & 31.00 & 28.18 & 22.00 & 18.46 & 21.00 & 0\\
			& $A_9B_3$ & 13.75 & 16  & 11.50 & 12.76 & 14.00 & 9.35 & 8.00 & 3.29 & 3.00  & 0\\
			& $A_{10}B_2$ & 4.13 & 0  & 3.50 & 3.52 & 3.50 & 1.92 & 2.00 & 0.34 & 0.50 & 0\\
			& $A_{11}B_1$ & 0.75 & 0  & 0.50 & 0.58 & 0.00 & 0.22 & 0.50 & 0.02 & 0.00 & 0\\
			& $A_{12}B_0$ & 0.06 & 0 & 0.00 & 0.04 & 0.00 & 0.01 & 0.00 & 0.00 & 0.00 & 0\\
		\end{tabular}
	\end{ruledtabular}
\end{table*}

Taking \ce{ZnSnP2} as an example, the ordered structure adopts a chalcopyrite configuration on a zinc-blende lattice, where the HOS is derived from hybridized P:$3p$ and Zn:$3d$ states, and the LUS arises from hybridized Sn:$5s$ and P:$3s$ state. To quantify the influence of local motifs, we construct ordered supercells embedding specific motifs and compute the evolution of the HOS and LUS states. 
As shown in Fig. \ref{fig:fig2}(c), when the anion-centered nearest-neighbor tetrahedron changes from \ce{Zn2Sn2} to \ce{Zn4Sn0{/}Zn0Sn4}, the HOS state rises while the LUS state falls, reducing the $E_g^{\text{LUS-HOS}}$ from 1.64 eV to 0.71 eV. This arises because increasing Zn atom near P strengthens the Zn:$3d$-P:$3p$ level repulsion, raising the HOS, while increasing Sn atom lowers the LUS due to the deeper Sn:$5s$ orbital and longer Sn-P bond length. Similarly, when fixing the nearest-neighbor tetrahedron as \ce{Zn4Sn0{/}Zn0Sn4}, and varying the second-nearest-neighbor polyhedron from Zn6Sn6 to \ce{Zn12Sn0{/}Zn0Sn12}, Sn-rich environments enhances the Sn-P contribution, further lowering the LUS state. In Zn-rich case, the HOS state and the LUS states crosses, confirming that minority motifs introduce defect-like states that can close the bandgap. More details on the structural models of polyhedra are provided in the Supplemental Materials (SM).

This explains why in disordered alloys, which statistically can include extreme minority motifs, exhibit significant reduction of $E_g^{\text{LUS-HOS}}$ and strong wave function localization near the band edges, leading to prominent band tails. Therefore, to correctly describe the bandgap of the disordered alloys, a new computational framework is needed to capture the electronic properties of disordered alloys by emphasizing majority configurations and excluding the artificial influence of minority motifs, as measured in experimental study.

\begin{figure*}[!t]
	\includegraphics[width=16cm]{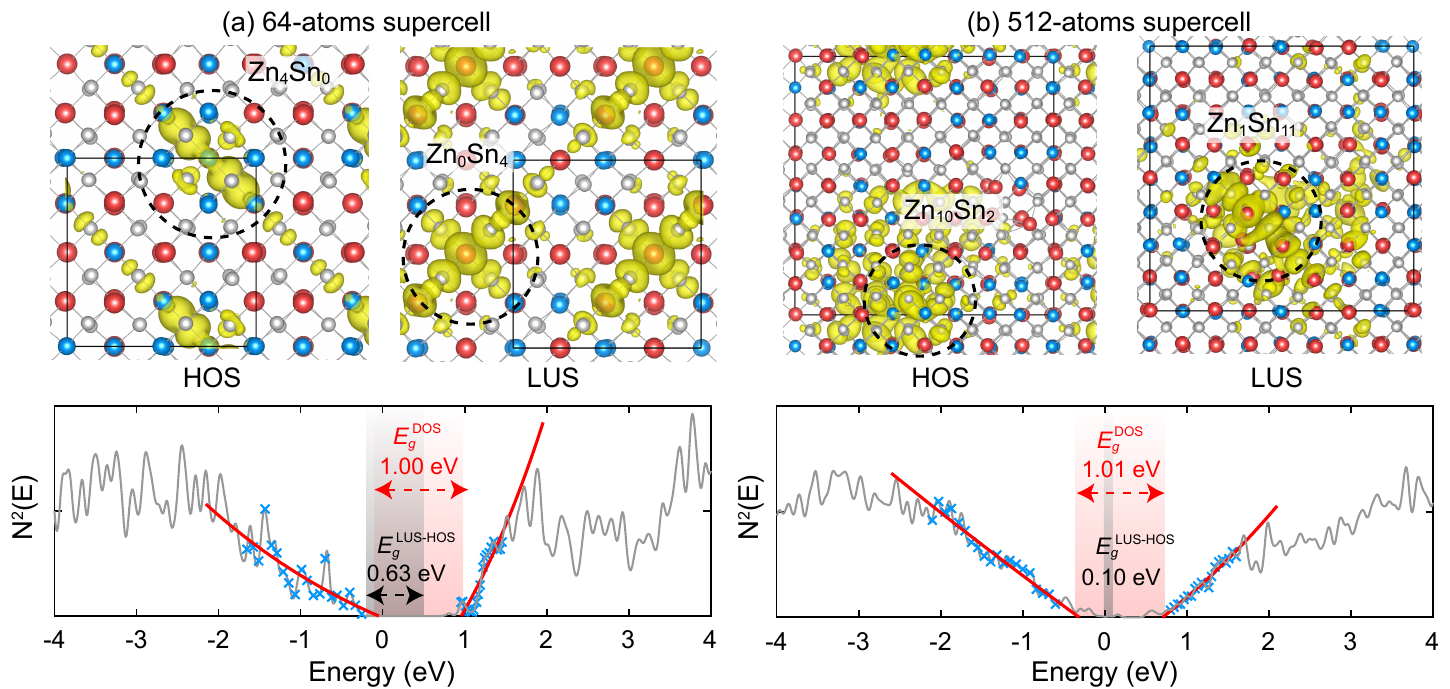}
	\caption{\label{fig:fig3} Density of states (DOS) for (a) 64-atom and (b) 512-atom supercells of disordered \ce{Zn_{0.5}Sn_{0.5}P} alloys. Blue crosses indicate the selected fitting points, while the red line represents the fitting trend. Charge distributions at the highest occupied state (HOS) and the lowest unoccupied state (LUS) are shown above the respective DOS panels.}
\end{figure*}

An electron in a parabolic band follows a general dispersion relation $E=\hbar^2 k^2/2m^*$, where $\hbar$ is the reduced Planck's constant, $k$  is the wave vector, and $m^*$ is the effective mass. The dependence of energy on $k^2$ implies that the DOS follows a $\sqrt{E}$ behavior, and thus the analytical expression for the DOS of a parabolic band can be derived as $N(E)=(2V/4\pi^2) \left(2m^*/\hbar^2\right)^{3/2} E^{1/2}$ or $N^2(E) \propto E$. By performing a linear fit at the band edge and extrapolating, an estimate of the bandgap can be obtained. However, in real physical systems, the band structure of a semiconductor compound is rarely parabolic due to the complex orbital coupling. To take into account the nonparabolicity, the dispersion relationship for a real semiconductor material can be effectively described by the function with energy dependent effective mass $m_d^*$, that is \cite{2000-JVTASF-band}
\begin{equation}
	\frac{\hbar^2k^2}{2m_d^*} = \gamma(E) = E + c_1 E^2 + c_2E^3 + \dots,
\end{equation}
where $c_1$ and $c_2$ are the nonparabolic coefficients. The relationship between $N(E)$ and $\gamma(E)$  is given by $N(E) \propto \gamma^{1/2} \left(\text{d}\gamma/\text{d}E\right)$. Thus, the DOS with the first-order energy correction term can be expressed as
\begin{equation}
	N(E)  \propto (E+c_1E^2)^{1/2}(1+2c_1E),
\end{equation}
thereby,
\begin{equation}
	N^2(E)  \propto E + 5c_1E^2 + 8c_1^2E^3+4c_1^3E^4.
\end{equation}
If we consider the second-order correction term, the square of the DOS is given by $N^2(E)  \propto  (E+c_1E^2+c_2E^3)(1+2c_1E+3c_2E^2)^2$  . By fitting the DOS near the Fermi level for both the valence band and conduction band, and determining the energies where $N_c^2 (E_{c,0} )=0$ and  $N_v^2 (E_{v,0} )=0$, the DOS bandgap can be determined as $E_{\text{g,DOS}}=E_{c,0}-E_{v,0}$. In disordered alloys, fitting the band edges can reveal the shape of the true band edges while filtering out the defect-like localized states induced by low-probability motifs. This DOSF approach, thus, provides an accurate determination of the bandgap for disordered alloys.

To test this method, we model the disordered alloys using the standard SQS \cite{1990-PRB-SQS,1990-PRL-SQS,atat-sqs} or special quasidisorder structure (SQDS) approaches \cite{1992-PRL-LRO,2009-PRB-LRO} . These approaches employ relatively small unit cells with periodic boundary conditions, where the mixed lattice sites are occupied by different atomic species to match the averaged atomic correlation functions of targeted alloys. Atomic relaxations and charge fluctuations needed to describe real alloys are naturally included in this approach. For ideal random alloys, the atomic correlation function for each cluster is given by $\langle \bar{\Pi}_R \rangle = (2x-1)^k$, where $x$ represents the alloy composition and $k$ is the number of vertices in the cluster. The structural relaxation are performed using density functional theory \cite{DFT-1,DFT-2} as implemented in the VASP \cite{VASP-1,VASP-2,VASP-3}. For accurate geometries and electronic structure calculations, we use the SCAN functional \cite{2015-PRL-SCAN,2018-PRM-SCAN} and the HSE06 functional \cite{HSE06}, respectively. 
More details on the SQS and SQDS, computational methods, and structural information are provided in the SM.

In theoretical alloy calculations, the size of the simulating supercell is an essential factor that needs to be considered. In principle, the atomic correlation function of a SQS supercell can match a fully random alloy better when the supercell size increases. 
In a simulation with 64-atom supercell containing 32 anions, it can effectively match the random distribution of nearest-neighbor tetrahedra, but it is impossible to match the correct distribution of long-range polyhedra because the number of available long-range polyhedra in a 64-atom supercell is inherently too small. 
However, in fully random alloys, the band edge states are strongly influenced by the appearance of large rare-event defect-like clusters. When the number of atoms in the supercell increases to 512, a more accurate distribution of long-range polyhedra such as $A_1B_{11}$ begin to appear, as shown in Tab. \ref{tab:tab1}. This leads to strong wave function localization at the band edge states, ultimately reducing the $E_g^{\text{LUS-HOS}}$.

\begin{table}[!t]
	\caption{Fitted nonparabolic coefficients $c_1$ and $c_2$ at the band edges of valence bands and conduction bands in 64-atom and 512-atom supercells of disordered \ce{Zn_{0.5}Sn_{0.5}P}.}\label{tab:tab2}
	\begin{ruledtabular}
		\renewcommand{\arraystretch}{1.4}
		\begin{tabular}{lccc}
			$N_{\text{atom}}$ & Band & $c_1$ (eV$^{-1}$) & $c_2$ ($10^{-7}$ eV$^{-2}$) \\
			\hline
			\multirow{2}{*}{64} & VB & 0.04 & 0.43 \\
			& CB & 0.06 & 0.70 \\
			\multirow{2}{*}{512} & VB & 0.10 & 0.10 \\
			& CB & 0.04 & 0.13 \\
		\end{tabular}
	\end{ruledtabular}
\end{table}

Taking \ce{Zn_{0.5}Sn_{0.5}P} as an example, Fig. \ref{fig:fig3} presents the DOS and its charge distribution at the band edges for these supercells. The calculated $E_g^{\text{LUS-HOS}}$ for the two supercells are 0.63 eV and 0.10 eV, respectively, with the former value being close to the reported calculation of 0.75 eV in the literature \cite{2012-APL-ZnSnP2-disorder,2014-PRB-ZnSnP2}. 
By defining the bandgap using a fit to the DOS at the band edges, that is, the DOS bandgap $E_{\text{g}}^{\text{DOS}}$, localized tail states within the gap can be effectively filtered out. Under this definition, the bandgaps for the two supercells are 1.00 eV and 1.01 eV. Table \ref{tab:tab2} lists the nonparabolic coefficients for 64-atom and 512-atom cubic supercells. These coefficients indicate that the band edge states of this disordered alloy system have indeed significant first-order nonparabolic characters ($c_1$), but the second-order term ($c_2$) is negligible, six orders of magnitude smaller than $c_1$.

\begin{figure}[!t]
	\includegraphics[width=8.5cm]{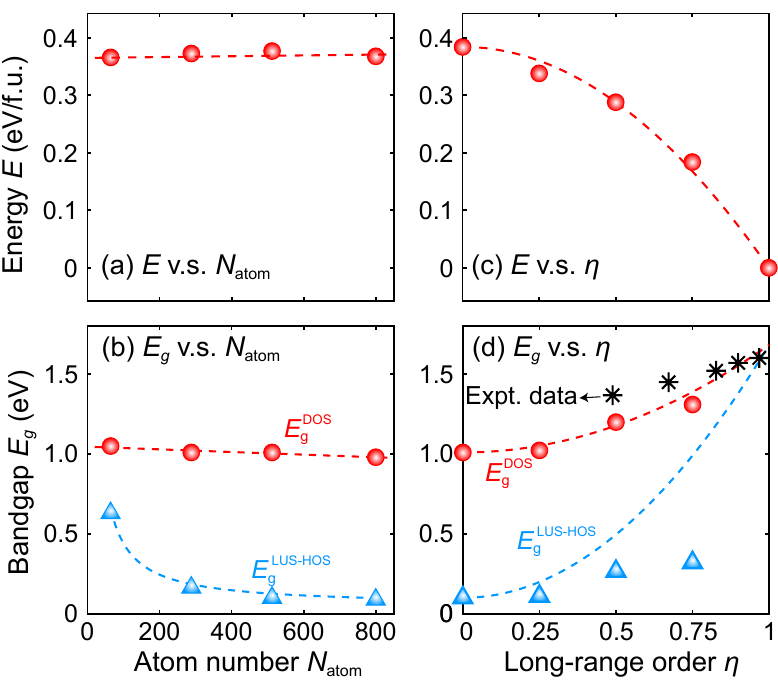}
	\caption{\label{fig:fig4} (a) Total energy $E$ relative to the order structure and (b) bandgap $E_g^{\text{LUS-HOS}}$ and $E_{\text{g}}^{\text{DOS}}$ as functions of atom number $N_{\text{atom}}$ in random \ce{Zn_{0.5}Sn_{0.5}P} supercells. (c) $E$ and (d) $E_g^{\text{LUS-HOS}}$ and $E_{\text{g}}^{\text{DOS}}$ as functions of long-range order $\eta$ in 512-atom supercells. Black star markers denote the experimental bandgaps of \ce{Zn_{0.5}Sn_{0.5}P} from Ref. \cite{2017-JPCC-ZnSnP2-expt-LRO}.}
\end{figure}

The differences in $E_g^{\text{LUS-HOS}}$ between the two supercell sizes arise from variations in the polyhedral distributions within each supercell. In the 64-atom supercell, the distribution of nearest tetrahedra closely matches that of an ideal random alloy. 
However, due to size limitations, the distribution of second-nearest-neighbor polyhedra is incomplete, with only configurations like \ce{Zn9Sn3} and \ce{Zn3Sn9} present. As a result, the band edge states are predominantly determined by the nearest-neighbor environments. 
The combined effect of Zn:$3d$ and Sn:$5s$ orbitals near the band edges leads to a reduced bandgap, with Zn-rich tetrahedra pushing up the valence band and Sn-rich tetrahedra lowering the conduction band. When the supercell size is increased to 512 atoms, second-nearest neighbor polyhedra such as \ce{Zn_1Sn_{11}} and \ce{Zn_{10}Sn_2} appear (see Tab. \ref{tab:tab1} and Fig. \ref{fig:fig3}), resulting in more localization of valence band edge state at the Zn-rich region and conduction band edge state at the Sn-rich region, so the $E_g^{\text{LUS-HOS}}$ become smaller. Figure \ref{fig:fig4}(a) and (b) illustrate the variation of the total energy and the two types of bandgaps with respect to supercell size. 
As the atom number in the supercell increases, the total energy converges to a stable value of around 0.38 eV/f.u. with respect to the ordered structure, primarily due to its dependence on the averaged atomic correlation functions, which remain consistent across different supercell sizes. In contrast, for the bandgap, the $E_g^{\text{LUS-HOS}}$ decreases progressively, while the $E_{\text{g}}^{\text{DOS}}$ converges to be around 1.0 eV for random alloy, closely aligning with the experimental estimate of 1.2 eV \cite{1987-JMR-ZnSnP2-expt,1999-APL-ZnSnP2-expt} for disordered alloy but much higher than previously calculated values between 0 to 0.7 eV, as seen in Fig. \ref{fig:fig1}(a) \cite{2012-APL-ZnSnP2-disorder,2014-PRB-ZnSnP2,2015-PRB-unconvergence}.

In real experimental situation at finite temperature, nonisovalent \ce{Zn_{0.5}Sn_{0.5}P} alloys is not fully random due to its high energy cost. Instead, they are only partially disordered at the order-disorder transition temperature \cite{1987-JMR-ZnSnP2-expt,2012-APL-ZnSnP2-disorder,2017-JPCC-ZnSnP2-expt-LRO}. In order to have a better comparison with experiment, we show here how our approach can be used to described partially disordered alloys. The long-range order parameter, denoted as $\eta$, serves as a parameter for quantifying the disorder in alloys \cite{1992-PRL-LRO,2009-PRB-LRO}. It also provides a straightforward way to interpolate experimental or theoretical data between perfectly random alloys ($\eta=0$) and fully ordered compounds ($\eta=1$). In most cases, the pair correlation is dominant, therefore, the atomic correlation function of pairs for partially disordered alloy can be written as 
\begin{equation}
	\bar{\Pi}_F(x, \eta) = (2x-1)^2 + \eta^2 [ \bar{\Pi}_F(\sigma)-(2X_\sigma-1)^2],
\end{equation}
where $X_\sigma$ is the composition of the fully ordered configuration $\sigma$ and $x$ is the alloy composition, $\bar{\Pi}_F(\sigma)$ is the atomic correlation function of the ordered compounds. The physical property that depends on $\eta$ can then be expressed as
\begin{equation}
	P(x, \eta) = P(x,0) + \eta^2[P(X_\sigma,1) - P(X_\sigma,0)].
\end{equation}
For \ce{Zn_{0.5}Sn_{0.5}P}, it is most stable in the chalcopyrite structure at the low temperature so this structure is chosen as the ordered configuration. 

By using the SQDS method to match the target atomic correlation function, we can generate supercells that represent partially disordered alloys. The atomic correlation functions and the DOS of the calculated partial disordered structure are provided in the SM. Table \ref{tab:tab1} lists the relative distribution of polyhedra at different long-range order $\eta$. In the fully ordered compound, the polyhedra are $A_2B_2$ and $A_6B_6$ only, respectively. As $\eta$ decreases, some atomic occupations start to deviate from their original ones, introducing octet-breaking polyhedra, which leads to the appearance of localized states at the band edges. As $\eta$ further decreases, the distribution of polyhedra gradually approaches that of a perfectly random alloy. 

Figure \ref{fig:fig4}(c) and (d) show the variation of the total energies and bandgaps with $\eta$ in 512-atom supercell with available experimental data from the literature \cite{2017-JPCC-ZnSnP2-expt-LRO}. Since the total energy is primarily influenced by averaged pair interactions, the $\eta^2$ rule for disordered alloys between ordered and random states is mostly obeyed. For the electronic structure, the induced localized states from the octet-breaking polyhedra significantly reduce the $E_g^{\text{LUS-HOS}}$ to nearly zero and deviate significantly from the parabolic trend. By filtering out the defect-like localized states within the band gap and fitting the DOS of the partial disordered structure, the $E_{\text{g}}^{\text{DOS}}$ follows the parabolic bandgap trend between the ordered configuration and random configurations, bringing it closer to the experimental observation. Compared to experimental values, at $\eta=0.5$, our calculated DOS bandgap is 1.22 eV, which is in reasonably good agreement with the experimentally observed bandgap of 1.39 eV \cite{2017-JPCC-ZnSnP2-expt-LRO}. 
This demonstrates that our approach can effectively address the challenge of the bandgap calculation in partially disordered and random alloys. 

In summary, to address the long-standing controversy in determining the non-statistical quantities, such as bandgap of the disordered alloys, resolving the disagreement between the calculated "bandgap" and experimental observations, we unveil that the poor convergence of $E_g^{\text{LUS-HOS}}$ with increasing supercell size in disordered alloy calculations arise from band edge wavefunction localization on minority configurations, whereas experimentally, the measured bandgap is $E_{\text{g}}^{\text{DOS}}$, which can be obtained by fitting the calculated electronic density of states to a well-established DOS function. $E_{\text{g}}^{\text{DOS}}$ and $E_g^{\text{LUS-HOS}}$ are the same for ordered compound as in conventional definition, but it can effectively filtering out the localized states induced by low-probability motifs in disordered alloys. Using \ce{Zn_{0.5}Sn_{0.5}P} as a case study, we demonstrate that the $E_{\text{g}}^{\text{DOS}}$ of fully disordered \ce{Zn_{0.5}Sn_{0.5}P} quickly converges to 1.0 eV, whereas $E_g^{\text{LUS-HOS}}$ does not converge and approach zero when the cell size increase. In partially disordered alloys, we find that the $E_{\text{g}}^{\text{DOS}}$ follow an $\eta^2$ rule with long-range order parameter $\eta$ and show good agreement with experimental observations. This work, thus, represents a significant advancement in the computational methods for studying electronic structures of disordered alloys.

\emph{Data availability} --- Computational code and examples are available in GitHub \cite{Github-code}.

\emph{Acknowledgments} --- This work was supported by the National Key Research and Development Program of China (Grants No. 2024YFA1409800).


\begin{thebibliography}{39}%
	\makeatletter
	\providecommand \@ifxundefined [1]{%
		\@ifx{#1\undefined}
	}%
	\providecommand \@ifnum [1]{%
		\ifnum #1\expandafter \@firstoftwo
		\else \expandafter \@secondoftwo
		\fi
	}%
	\providecommand \@ifx [1]{%
		\ifx #1\expandafter \@firstoftwo
		\else \expandafter \@secondoftwo
		\fi
	}%
	\providecommand \natexlab [1]{#1}%
	\providecommand \enquote  [1]{``#1''}%
	\providecommand \bibnamefont  [1]{#1}%
	\providecommand \bibfnamefont [1]{#1}%
	\providecommand \citenamefont [1]{#1}%
	\providecommand \href@noop [0]{\@secondoftwo}%
	\providecommand \href [0]{\begingroup \@sanitize@url \@href}%
	\providecommand \@href[1]{\@@startlink{#1}\@@href}%
	\providecommand \@@href[1]{\endgroup#1\@@endlink}%
	\providecommand \@sanitize@url [0]{\catcode `\\12\catcode `\$12\catcode
		`\&12\catcode `\#12\catcode `\^12\catcode `\_12\catcode `\%12\relax}%
	\providecommand \@@startlink[1]{}%
	\providecommand \@@endlink[0]{}%
	\providecommand \url  [0]{\begingroup\@sanitize@url \@url }%
	\providecommand \@url [1]{\endgroup\@href {#1}{\urlprefix }}%
	\providecommand \urlprefix  [0]{URL }%
	\providecommand \Eprint [0]{\href }%
	\providecommand \doibase [0]{http://dx.doi.org/}%
	\providecommand \selectlanguage [0]{\@gobble}%
	\providecommand \bibinfo  [0]{\@secondoftwo}%
	\providecommand \bibfield  [0]{\@secondoftwo}%
	\providecommand \translation [1]{[#1]}%
	\providecommand \BibitemOpen [0]{}%
	\providecommand \bibitemStop [0]{}%
	\providecommand \bibitemNoStop [0]{.\EOS\space}%
	\providecommand \EOS [0]{\spacefactor3000\relax}%
	\providecommand \BibitemShut  [1]{\csname bibitem#1\endcsname}%
	\let\auto@bib@innerbib\@empty
	\bibitem [{\citenamefont {Zhang}\ \emph {et~al.}(2008)\citenamefont {Zhang},
		\citenamefont {Mascarenhas},\ and\ \citenamefont
		{Wang}}]{2008-PRL-GaInP-alloy}%
	\BibitemOpen
	\bibfield  {author} {\bibinfo {author} {\bibfnamefont {Y.}~\bibnamefont
			{Zhang}}, \bibinfo {author} {\bibfnamefont {A.}~\bibnamefont {Mascarenhas}},
		\ and\ \bibinfo {author} {\bibfnamefont {L.-W.}\ \bibnamefont {Wang}},\
	}\href {\doibase 10.1103/PhysRevLett.101.036403} {\bibfield  {journal}
		{\bibinfo  {journal} {Phys. Rev. Lett.}\ }\textbf {\bibinfo {volume} {101}},\
		\bibinfo {pages} {036403} (\bibinfo {year} {2008})}\BibitemShut {NoStop}%
	\bibitem [{\citenamefont {Voas}\ \emph {et~al.}(2014)\citenamefont {Voas},
		\citenamefont {Usher}, \citenamefont {Liu}, \citenamefont {Li}, \citenamefont
		{Jones}, \citenamefont {Tan}, \citenamefont {Cooper},\ and\ \citenamefont
		{Beckman}}]{2014-PRB-perovskite}%
	\BibitemOpen
	\bibfield  {author} {\bibinfo {author} {\bibfnamefont {B.~K.}\ \bibnamefont
			{Voas}}, \bibinfo {author} {\bibfnamefont {T.-M.}\ \bibnamefont {Usher}},
		\bibinfo {author} {\bibfnamefont {X.}~\bibnamefont {Liu}}, \bibinfo {author}
		{\bibfnamefont {S.}~\bibnamefont {Li}}, \bibinfo {author} {\bibfnamefont
			{J.~L.}\ \bibnamefont {Jones}}, \bibinfo {author} {\bibfnamefont
			{X.}~\bibnamefont {Tan}}, \bibinfo {author} {\bibfnamefont {V.~R.}\
			\bibnamefont {Cooper}}, \ and\ \bibinfo {author} {\bibfnamefont {S.~P.}\
			\bibnamefont {Beckman}},\ }\href {\doibase 10.1103/PhysRevB.90.024105}
	{\bibfield  {journal} {\bibinfo  {journal} {Phys. Rev. B}\ }\textbf {\bibinfo
			{volume} {90}},\ \bibinfo {pages} {024105} (\bibinfo {year}
		{2014})}\BibitemShut {NoStop}%
	\bibitem [{\citenamefont {Sauty}\ \emph {et~al.}(2022)\citenamefont {Sauty},
		\citenamefont {Lopes}, \citenamefont {Banon}, \citenamefont {Lassailly},
		\citenamefont {Martinelli}, \citenamefont {Alhassan}, \citenamefont {Chow},
		\citenamefont {Nakamura}, \citenamefont {Speck}, \citenamefont {Weisbuch},\
		and\ \citenamefont {Peretti}}]{2022-PRL-Nitride-alloy}%
	\BibitemOpen
	\bibfield  {author} {\bibinfo {author} {\bibfnamefont {M.}~\bibnamefont
			{Sauty}}, \bibinfo {author} {\bibfnamefont {N.~M.~S.}\ \bibnamefont {Lopes}},
		\bibinfo {author} {\bibfnamefont {J.-P.}\ \bibnamefont {Banon}}, \bibinfo
		{author} {\bibfnamefont {Y.}~\bibnamefont {Lassailly}}, \bibinfo {author}
		{\bibfnamefont {L.}~\bibnamefont {Martinelli}}, \bibinfo {author}
		{\bibfnamefont {A.}~\bibnamefont {Alhassan}}, \bibinfo {author}
		{\bibfnamefont {Y.~C.}\ \bibnamefont {Chow}}, \bibinfo {author}
		{\bibfnamefont {S.}~\bibnamefont {Nakamura}}, \bibinfo {author}
		{\bibfnamefont {J.~S.}\ \bibnamefont {Speck}}, \bibinfo {author}
		{\bibfnamefont {C.}~\bibnamefont {Weisbuch}}, \ and\ \bibinfo {author}
		{\bibfnamefont {J.}~\bibnamefont {Peretti}},\ }\href {\doibase
		10.1103/PhysRevLett.129.216602} {\bibfield  {journal} {\bibinfo  {journal}
			{Phys. Rev. Lett.}\ }\textbf {\bibinfo {volume} {129}},\ \bibinfo {pages}
		{216602} (\bibinfo {year} {2022})}\BibitemShut {NoStop}%
	\bibitem [{\citenamefont {Wang}\ \emph {et~al.}(2022)\citenamefont {Wang},
		\citenamefont {Kavanagh}, \citenamefont {Burgu\'es-Ceballos}, \citenamefont
		{Walsh}, \citenamefont {Scanlon},\ and\ \citenamefont
		{Konstantatos}}]{2022-NP-AgBiS2}%
	\BibitemOpen
	\bibfield  {author} {\bibinfo {author} {\bibfnamefont {Y.}~\bibnamefont
			{Wang}}, \bibinfo {author} {\bibfnamefont {S.~R.}\ \bibnamefont {Kavanagh}},
		\bibinfo {author} {\bibfnamefont {I.}~\bibnamefont {Burgu\'es-Ceballos}},
		\bibinfo {author} {\bibfnamefont {A.}~\bibnamefont {Walsh}}, \bibinfo
		{author} {\bibfnamefont {D.~O.}\ \bibnamefont {Scanlon}}, \ and\ \bibinfo
		{author} {\bibfnamefont {G.}~\bibnamefont {Konstantatos}},\ }\href {\doibase
		10.1038/s41566-021-00950-4} {\bibfield  {journal} {\bibinfo  {journal} {Nat.
				Photon.}\ }\textbf {\bibinfo {volume} {16}},\ \bibinfo {pages} {235}
		(\bibinfo {year} {2022})}\BibitemShut {NoStop}%
	\bibitem [{\citenamefont {Zhang}\ \emph {et~al.}(2023)\citenamefont {Zhang},
		\citenamefont {Kang},\ and\ \citenamefont
		{Wei}}]{2023-NCS-defect-alloy-review}%
	\BibitemOpen
	\bibfield  {author} {\bibinfo {author} {\bibfnamefont {X.}~\bibnamefont
			{Zhang}}, \bibinfo {author} {\bibfnamefont {J.}~\bibnamefont {Kang}}, \ and\
		\bibinfo {author} {\bibfnamefont {S.-H.}\ \bibnamefont {Wei}},\ }\href
	{\doibase 10.1038/s43588-023-00403-8} {\bibfield  {journal} {\bibinfo
			{journal} {Nat. Comput. Sci.}\ }\textbf {\bibinfo {volume} {3}},\ \bibinfo
		{pages} {210} (\bibinfo {year} {2023})}\BibitemShut {NoStop}%
	\bibitem [{\citenamefont {Wang}\ \emph {et~al.}(2024)\citenamefont {Wang},
		\citenamefont {Yao}, \citenamefont {Wang}, \citenamefont {Guo}, \citenamefont
		{Li}, \citenamefont {Zhou}, \citenamefont {Bai}, \citenamefont {Li},
		\citenamefont {Li}, \citenamefont {Wagemaker},\ and\ \citenamefont
		{Zhao}}]{2024-Nature-SQS}%
	\BibitemOpen
	\bibfield  {author} {\bibinfo {author} {\bibfnamefont {Q.}~\bibnamefont
			{Wang}}, \bibinfo {author} {\bibfnamefont {Z.}~\bibnamefont {Yao}}, \bibinfo
		{author} {\bibfnamefont {J.}~\bibnamefont {Wang}}, \bibinfo {author}
		{\bibfnamefont {H.}~\bibnamefont {Guo}}, \bibinfo {author} {\bibfnamefont
			{C.}~\bibnamefont {Li}}, \bibinfo {author} {\bibfnamefont {D.}~\bibnamefont
			{Zhou}}, \bibinfo {author} {\bibfnamefont {X.}~\bibnamefont {Bai}}, \bibinfo
		{author} {\bibfnamefont {H.}~\bibnamefont {Li}}, \bibinfo {author}
		{\bibfnamefont {B.}~\bibnamefont {Li}}, \bibinfo {author} {\bibfnamefont
			{M.}~\bibnamefont {Wagemaker}}, \ and\ \bibinfo {author} {\bibfnamefont
			{C.}~\bibnamefont {Zhao}},\ }\href {\doibase 10.1038/s41586-024-07362-8}
	{\bibfield  {journal} {\bibinfo  {journal} {Nature}\ }\textbf {\bibinfo
			{volume} {629}},\ \bibinfo {pages} {341} (\bibinfo {year}
		{2024})}\BibitemShut {NoStop}%
	\bibitem [{\citenamefont {Huang}\ \emph {et~al.}(2024)\citenamefont {Huang},
		\citenamefont {Li}, \citenamefont {Hao}, \citenamefont {Liang}, \citenamefont
		{Cai}, \citenamefont {Yue}, \citenamefont {Kuznetsov}, \citenamefont
		{Zhang},\ and\ \citenamefont {Wei}}]{2024-PRL-disorder-Ga2O3}%
	\BibitemOpen
	\bibfield  {author} {\bibinfo {author} {\bibfnamefont {Q.-S.}\ \bibnamefont
			{Huang}}, \bibinfo {author} {\bibfnamefont {C.-N.}\ \bibnamefont {Li}},
		\bibinfo {author} {\bibfnamefont {M.-S.}\ \bibnamefont {Hao}}, \bibinfo
		{author} {\bibfnamefont {H.-P.}\ \bibnamefont {Liang}}, \bibinfo {author}
		{\bibfnamefont {X.}~\bibnamefont {Cai}}, \bibinfo {author} {\bibfnamefont
			{Y.}~\bibnamefont {Yue}}, \bibinfo {author} {\bibfnamefont {A.}~\bibnamefont
			{Kuznetsov}}, \bibinfo {author} {\bibfnamefont {X.}~\bibnamefont {Zhang}}, \
		and\ \bibinfo {author} {\bibfnamefont {S.-H.}\ \bibnamefont {Wei}},\ }\href
	{\doibase 10.1103/PhysRevLett.133.226101} {\bibfield  {journal} {\bibinfo
			{journal} {Phys. Rev. Lett.}\ }\textbf {\bibinfo {volume} {133}},\ \bibinfo
		{pages} {226101} (\bibinfo {year} {2024})}\BibitemShut {NoStop}%
	\bibitem [{\citenamefont {Wei}\ \emph {et~al.}(1990)\citenamefont {Wei},
		\citenamefont {Ferreira}, \citenamefont {Bernard},\ and\ \citenamefont
		{Zunger}}]{1990-PRB-SQS}%
	\BibitemOpen
	\bibfield  {author} {\bibinfo {author} {\bibfnamefont {S.-H.}\ \bibnamefont
			{Wei}}, \bibinfo {author} {\bibfnamefont {L.~G.}\ \bibnamefont {Ferreira}},
		\bibinfo {author} {\bibfnamefont {J.~E.}\ \bibnamefont {Bernard}}, \ and\
		\bibinfo {author} {\bibfnamefont {A.}~\bibnamefont {Zunger}},\ }\href
	{\doibase 10.1103/PhysRevB.42.9622} {\bibfield  {journal} {\bibinfo
			{journal} {Phys. Rev. B}\ }\textbf {\bibinfo {volume} {42}},\ \bibinfo
		{pages} {9622} (\bibinfo {year} {1990})}\BibitemShut {NoStop}%
	\bibitem [{\citenamefont {Zunger}\ \emph {et~al.}(1990)\citenamefont {Zunger},
		\citenamefont {Wei}, \citenamefont {Ferreira},\ and\ \citenamefont
		{Bernard}}]{1990-PRL-SQS}%
	\BibitemOpen
	\bibfield  {author} {\bibinfo {author} {\bibfnamefont {A.}~\bibnamefont
			{Zunger}}, \bibinfo {author} {\bibfnamefont {S.-H.}\ \bibnamefont {Wei}},
		\bibinfo {author} {\bibfnamefont {L.~G.}\ \bibnamefont {Ferreira}}, \ and\
		\bibinfo {author} {\bibfnamefont {J.~E.}\ \bibnamefont {Bernard}},\ }\href
	{\doibase 10.1103/PhysRevLett.65.353} {\bibfield  {journal} {\bibinfo
			{journal} {Phys. Rev. Lett.}\ }\textbf {\bibinfo {volume} {65}},\ \bibinfo
		{pages} {353} (\bibinfo {year} {1990})}\BibitemShut {NoStop}%
	\bibitem [{\citenamefont {{van de Walle}}\ \emph {et~al.}(2013)\citenamefont
		{{van de Walle}}, \citenamefont {Tiwary}, \citenamefont {{de Jong}},
		\citenamefont {Olmsted}, \citenamefont {Asta}, \citenamefont {Dick},
		\citenamefont {Shin}, \citenamefont {Wang}, \citenamefont {Chen},\ and\
		\citenamefont {Liu}}]{atat-sqs}%
	\BibitemOpen
	\bibfield  {author} {\bibinfo {author} {\bibfnamefont {A.}~\bibnamefont {{van
					de Walle}}}, \bibinfo {author} {\bibfnamefont {P.}~\bibnamefont {Tiwary}},
		\bibinfo {author} {\bibfnamefont {M.}~\bibnamefont {{de Jong}}}, \bibinfo
		{author} {\bibfnamefont {D.}~\bibnamefont {Olmsted}}, \bibinfo {author}
		{\bibfnamefont {M.}~\bibnamefont {Asta}}, \bibinfo {author} {\bibfnamefont
			{A.}~\bibnamefont {Dick}}, \bibinfo {author} {\bibfnamefont {D.}~\bibnamefont
			{Shin}}, \bibinfo {author} {\bibfnamefont {Y.}~\bibnamefont {Wang}}, \bibinfo
		{author} {\bibfnamefont {L.-Q.}\ \bibnamefont {Chen}}, \ and\ \bibinfo
		{author} {\bibfnamefont {Z.-K.}\ \bibnamefont {Liu}},\ }\href {\doibase
		https://doi.org/10.1016/j.calphad.2013.06.006} {\bibfield  {journal}
		{\bibinfo  {journal} {Calphad}\ }\textbf {\bibinfo {volume} {42}},\ \bibinfo
		{pages} {13} (\bibinfo {year} {2013})}\BibitemShut {NoStop}%
	\bibitem [{\citenamefont {Laks}\ \emph {et~al.}(1992)\citenamefont {Laks},
		\citenamefont {Wei},\ and\ \citenamefont {Zunger}}]{1992-PRL-LRO}%
	\BibitemOpen
	\bibfield  {author} {\bibinfo {author} {\bibfnamefont {D.~B.}\ \bibnamefont
			{Laks}}, \bibinfo {author} {\bibfnamefont {S.-H.}\ \bibnamefont {Wei}}, \
		and\ \bibinfo {author} {\bibfnamefont {A.}~\bibnamefont {Zunger}},\ }\href
	{\doibase 10.1103/PhysRevLett.69.3766} {\bibfield  {journal} {\bibinfo
			{journal} {Phys. Rev. Lett.}\ }\textbf {\bibinfo {volume} {69}},\ \bibinfo
		{pages} {3766} (\bibinfo {year} {1992})}\BibitemShut {NoStop}%
	\bibitem [{\citenamefont {Wolverton}\ and\ \citenamefont
		{Zunger}(1998)}]{1998-PRL-LiCoO2}%
	\BibitemOpen
	\bibfield  {author} {\bibinfo {author} {\bibfnamefont {C.}~\bibnamefont
			{Wolverton}}\ and\ \bibinfo {author} {\bibfnamefont {A.}~\bibnamefont
			{Zunger}},\ }\href {\doibase 10.1103/PhysRevLett.81.606} {\bibfield
		{journal} {\bibinfo  {journal} {Phys. Rev. Lett.}\ }\textbf {\bibinfo
			{volume} {81}},\ \bibinfo {pages} {606} (\bibinfo {year} {1998})}\BibitemShut
	{NoStop}%
	\bibitem [{\citenamefont {Saitta}\ \emph {et~al.}(1998)\citenamefont {Saitta},
		\citenamefont {de~Gironcoli},\ and\ \citenamefont
		{Baroni}}]{1998-PRL-ZnMgSSe}%
	\BibitemOpen
	\bibfield  {author} {\bibinfo {author} {\bibfnamefont {A.~M.}\ \bibnamefont
			{Saitta}}, \bibinfo {author} {\bibfnamefont {S.}~\bibnamefont
			{de~Gironcoli}}, \ and\ \bibinfo {author} {\bibfnamefont {S.}~\bibnamefont
			{Baroni}},\ }\href {\doibase 10.1103/PhysRevLett.80.4939} {\bibfield
		{journal} {\bibinfo  {journal} {Phys. Rev. Lett.}\ }\textbf {\bibinfo
			{volume} {80}},\ \bibinfo {pages} {4939} (\bibinfo {year}
		{1998})}\BibitemShut {NoStop}%
	\bibitem [{\citenamefont {Segev}\ and\ \citenamefont
		{Wei}(2004)}]{2004-PRB-dilute}%
	\BibitemOpen
	\bibfield  {author} {\bibinfo {author} {\bibfnamefont {D.}~\bibnamefont
			{Segev}}\ and\ \bibinfo {author} {\bibfnamefont {S.-H.}\ \bibnamefont
			{Wei}},\ }\href {\doibase 10.1103/PhysRevB.70.184401} {\bibfield  {journal}
		{\bibinfo  {journal} {Phys. Rev. B}\ }\textbf {\bibinfo {volume} {70}},\
		\bibinfo {pages} {184401} (\bibinfo {year} {2004})}\BibitemShut {NoStop}%
	\bibitem [{\citenamefont {Liu}\ \emph {et~al.}(2019)\citenamefont {Liu},
		\citenamefont {Liang}, \citenamefont {Duan},\ and\ \citenamefont
		{Wu}}]{2019-PRM-CuGaSe2-disorder}%
	\BibitemOpen
	\bibfield  {author} {\bibinfo {author} {\bibfnamefont {W.}~\bibnamefont
			{Liu}}, \bibinfo {author} {\bibfnamefont {H.}~\bibnamefont {Liang}}, \bibinfo
		{author} {\bibfnamefont {Y.}~\bibnamefont {Duan}}, \ and\ \bibinfo {author}
		{\bibfnamefont {Z.}~\bibnamefont {Wu}},\ }\href {\doibase
		10.1103/PhysRevMaterials.3.125405} {\bibfield  {journal} {\bibinfo  {journal}
			{Phys. Rev. Mater.}\ }\textbf {\bibinfo {volume} {3}},\ \bibinfo {pages}
		{125405} (\bibinfo {year} {2019})}\BibitemShut {NoStop}%
	\bibitem [{\citenamefont {Liang}\ \emph {et~al.}(2024)\citenamefont {Liang},
		\citenamefont {Li}, \citenamefont {Zhou}, \citenamefont {Xu}, \citenamefont
		{Zhang}, \citenamefont {Yang},\ and\ \citenamefont {Wei}}]{2024-JACS-ABX2}%
	\BibitemOpen
	\bibfield  {author} {\bibinfo {author} {\bibfnamefont {H.-P.}\ \bibnamefont
			{Liang}}, \bibinfo {author} {\bibfnamefont {C.-N.}\ \bibnamefont {Li}},
		\bibinfo {author} {\bibfnamefont {R.}~\bibnamefont {Zhou}}, \bibinfo {author}
		{\bibfnamefont {X.}~\bibnamefont {Xu}}, \bibinfo {author} {\bibfnamefont
			{X.}~\bibnamefont {Zhang}}, \bibinfo {author} {\bibfnamefont
			{J.}~\bibnamefont {Yang}}, \ and\ \bibinfo {author} {\bibfnamefont {S.-H.}\
			\bibnamefont {Wei}},\ }\href {\doibase 10.1021/jacs.4c04201} {\bibfield
		{journal} {\bibinfo  {journal} {J. Am. Chem. Soc.}\ }\textbf {\bibinfo
			{volume} {146}},\ \bibinfo {pages} {16222} (\bibinfo {year}
		{2024})}\BibitemShut {NoStop}%
	\bibitem [{\citenamefont {Siegrist}\ \emph {et~al.}(2011)\citenamefont
		{Siegrist}, \citenamefont {Jost}, \citenamefont {Volker}, \citenamefont
		{Woda}, \citenamefont {Merkelbach}, \citenamefont {Schlockermann},\ and\
		\citenamefont {Wuttig}}]{2011-NM-disorderalloy}%
	\BibitemOpen
	\bibfield  {author} {\bibinfo {author} {\bibfnamefont {T.}~\bibnamefont
			{Siegrist}}, \bibinfo {author} {\bibfnamefont {P.}~\bibnamefont {Jost}},
		\bibinfo {author} {\bibfnamefont {H.}~\bibnamefont {Volker}}, \bibinfo
		{author} {\bibfnamefont {M.}~\bibnamefont {Woda}}, \bibinfo {author}
		{\bibfnamefont {P.}~\bibnamefont {Merkelbach}}, \bibinfo {author}
		{\bibfnamefont {C.}~\bibnamefont {Schlockermann}}, \ and\ \bibinfo {author}
		{\bibfnamefont {M.}~\bibnamefont {Wuttig}},\ }\href {\doibase
		10.1038/nmat2934} {\bibfield  {journal} {\bibinfo  {journal} {Nat. Mater.}\
		}\textbf {\bibinfo {volume} {10}},\ \bibinfo {pages} {202} (\bibinfo {year}
		{2011})}\BibitemShut {NoStop}%
	\bibitem [{\citenamefont {Larsen}\ \emph {et~al.}(2020)\citenamefont {Larsen},
		\citenamefont {Scragg}, \citenamefont {Ross},\ and\ \citenamefont
		{{Platzer-Bj{\"o}rkman}}}]{2020-ACS-bandtail}%
	\BibitemOpen
	\bibfield  {author} {\bibinfo {author} {\bibfnamefont {J.~K.}\ \bibnamefont
			{Larsen}}, \bibinfo {author} {\bibfnamefont {J.~J.~S.}\ \bibnamefont
			{Scragg}}, \bibinfo {author} {\bibfnamefont {N.}~\bibnamefont {Ross}}, \ and\
		\bibinfo {author} {\bibfnamefont {C.}~\bibnamefont
			{{Platzer-Bj{\"o}rkman}}},\ }\href {\doibase 10.1021/acsaem.0c00926}
	{\bibfield  {journal} {\bibinfo  {journal} {ACS Appl. Energy Mater.}\
		}\textbf {\bibinfo {volume} {3}},\ \bibinfo {pages} {7520} (\bibinfo {year}
		{2020})}\BibitemShut {NoStop}%
	\bibitem [{\citenamefont {Bleuse}\ \emph {et~al.}(2020)\citenamefont {Bleuse},
		\citenamefont {Perret}, \citenamefont {Cur\'e}, \citenamefont {Grenet},
		\citenamefont {Andr\'e},\ and\ \citenamefont {Mariette}}]{2020-PRB-bandtail}%
	\BibitemOpen
	\bibfield  {author} {\bibinfo {author} {\bibfnamefont {J.}~\bibnamefont
			{Bleuse}}, \bibinfo {author} {\bibfnamefont {S.}~\bibnamefont {Perret}},
		\bibinfo {author} {\bibfnamefont {Y.}~\bibnamefont {Cur\'e}}, \bibinfo
		{author} {\bibfnamefont {L.}~\bibnamefont {Grenet}}, \bibinfo {author}
		{\bibfnamefont {R.}~\bibnamefont {Andr\'e}}, \ and\ \bibinfo {author}
		{\bibfnamefont {H.}~\bibnamefont {Mariette}},\ }\href {\doibase
		10.1103/PhysRevB.102.195205} {\bibfield  {journal} {\bibinfo  {journal}
			{Phys. Rev. B}\ }\textbf {\bibinfo {volume} {102}},\ \bibinfo {pages}
		{195205} (\bibinfo {year} {2020})}\BibitemShut {NoStop}%
	\bibitem [{\citenamefont {Baranovskii}\ \emph {et~al.}(2022)\citenamefont
		{Baranovskii}, \citenamefont {Nenashev}, \citenamefont {Hertel},
		\citenamefont {Gebhard},\ and\ \citenamefont
		{Meerholz}}]{2022-ACSO-alloy-review}%
	\BibitemOpen
	\bibfield  {author} {\bibinfo {author} {\bibfnamefont {S.~D.}\ \bibnamefont
			{Baranovskii}}, \bibinfo {author} {\bibfnamefont {A.~V.}\ \bibnamefont
			{Nenashev}}, \bibinfo {author} {\bibfnamefont {D.}~\bibnamefont {Hertel}},
		\bibinfo {author} {\bibfnamefont {F.}~\bibnamefont {Gebhard}}, \ and\
		\bibinfo {author} {\bibfnamefont {K.}~\bibnamefont {Meerholz}},\ }\href
	{\doibase 10.1021/acsomega.2c05426} {\bibfield  {journal} {\bibinfo
			{journal} {ACS Omega}\ }\textbf {\bibinfo {volume} {7}},\ \bibinfo {pages}
		{45741} (\bibinfo {year} {2022})}\BibitemShut {NoStop}%
	\bibitem [{\citenamefont {Tauc}(1968)}]{1968-MRB-Tauc}%
	\BibitemOpen
	\bibfield  {author} {\bibinfo {author} {\bibfnamefont {J.}~\bibnamefont
			{Tauc}},\ }\href {\doibase https://doi.org/10.1016/0025-5408(68)90023-8}
	{\bibfield  {journal} {\bibinfo  {journal} {Mater. Res. Bull.}\ }\textbf
		{\bibinfo {volume} {3}},\ \bibinfo {pages} {37} (\bibinfo {year}
		{1968})}\BibitemShut {NoStop}%
	\bibitem [{\citenamefont {Ma}\ \emph {et~al.}(2014)\citenamefont {Ma},
		\citenamefont {Deng}, \citenamefont {Luo},\ and\ \citenamefont
		{Wei}}]{2014-PRB-ZnSnP2}%
	\BibitemOpen
	\bibfield  {author} {\bibinfo {author} {\bibfnamefont {J.}~\bibnamefont
			{Ma}}, \bibinfo {author} {\bibfnamefont {H.-X.}\ \bibnamefont {Deng}},
		\bibinfo {author} {\bibfnamefont {J.-W.}\ \bibnamefont {Luo}}, \ and\
		\bibinfo {author} {\bibfnamefont {S.-H.}\ \bibnamefont {Wei}},\ }\href
	{\doibase 10.1103/PhysRevB.90.115201} {\bibfield  {journal} {\bibinfo
			{journal} {Phys. Rev. B}\ }\textbf {\bibinfo {volume} {90}},\ \bibinfo
		{pages} {115201} (\bibinfo {year} {2014})}\BibitemShut {NoStop}%
	\bibitem [{\citenamefont {Scanlon}\ and\ \citenamefont
		{Walsh}(2012)}]{2012-APL-ZnSnP2-disorder}%
	\BibitemOpen
	\bibfield  {author} {\bibinfo {author} {\bibfnamefont {D.~O.}\ \bibnamefont
			{Scanlon}}\ and\ \bibinfo {author} {\bibfnamefont {A.}~\bibnamefont
			{Walsh}},\ }\href {\doibase 10.1063/1.4730375} {\bibfield  {journal}
		{\bibinfo  {journal} {Appl. Phys. Lett.}\ }\textbf {\bibinfo {volume}
			{100}},\ \bibinfo {pages} {251911} (\bibinfo {year} {2012})}\BibitemShut
	{NoStop}%
	\bibitem [{\citenamefont {Seko}\ and\ \citenamefont
		{Tanaka}(2015)}]{2015-PRB-unconvergence}%
	\BibitemOpen
	\bibfield  {author} {\bibinfo {author} {\bibfnamefont {A.}~\bibnamefont
			{Seko}}\ and\ \bibinfo {author} {\bibfnamefont {I.}~\bibnamefont {Tanaka}},\
	}\href {\doibase 10.1103/PhysRevB.91.024106} {\bibfield  {journal} {\bibinfo
			{journal} {Phys. Rev. B}\ }\textbf {\bibinfo {volume} {91}},\ \bibinfo
		{pages} {024106} (\bibinfo {year} {2015})}\BibitemShut {NoStop}%
	\bibitem [{\citenamefont {Nakatsuka}\ and\ \citenamefont
		{Nose}(2017)}]{2017-JPCC-ZnSnP2-expt-LRO}%
	\BibitemOpen
	\bibfield  {author} {\bibinfo {author} {\bibfnamefont {S.}~\bibnamefont
			{Nakatsuka}}\ and\ \bibinfo {author} {\bibfnamefont {Y.}~\bibnamefont
			{Nose}},\ }\href {\doibase 10.1021/acs.jpcc.6b11215} {\bibfield  {journal}
		{\bibinfo  {journal} {J. Phys. Chem. C}\ }\textbf {\bibinfo {volume} {121}},\
		\bibinfo {pages} {1040} (\bibinfo {year} {2017})}\BibitemShut {NoStop}%
	\bibitem [{\citenamefont {Ryan}\ \emph {et~al.}(1987)\citenamefont {Ryan},
		\citenamefont {Peterson}, \citenamefont {Williamson}, \citenamefont {Frey},
		\citenamefont {Maciel},\ and\ \citenamefont
		{Parkinson}}]{1987-JMR-ZnSnP2-expt}%
	\BibitemOpen
	\bibfield  {author} {\bibinfo {author} {\bibfnamefont {M.~A.}\ \bibnamefont
			{Ryan}}, \bibinfo {author} {\bibfnamefont {M.~W.}\ \bibnamefont {Peterson}},
		\bibinfo {author} {\bibfnamefont {D.~L.}\ \bibnamefont {Williamson}},
		\bibinfo {author} {\bibfnamefont {J.~S.}\ \bibnamefont {Frey}}, \bibinfo
		{author} {\bibfnamefont {G.~E.}\ \bibnamefont {Maciel}}, \ and\ \bibinfo
		{author} {\bibfnamefont {B.~A.}\ \bibnamefont {Parkinson}},\ }\href {\doibase
		10.1557/JMR.1987.0528} {\bibfield  {journal} {\bibinfo  {journal} {J. Mater.
				Res.}\ }\textbf {\bibinfo {volume} {2}},\ \bibinfo {pages} {528} (\bibinfo
		{year} {1987})}\BibitemShut {NoStop}%
	\bibitem [{\citenamefont {Nakatsuka}\ \emph {et~al.}(2015)\citenamefont
		{Nakatsuka}, \citenamefont {Nakamoto}, \citenamefont {Nose}, \citenamefont
		{Uda},\ and\ \citenamefont {Shirai}}]{2015-PSSC-ZnSnP2}%
	\BibitemOpen
	\bibfield  {author} {\bibinfo {author} {\bibfnamefont {S.}~\bibnamefont
			{Nakatsuka}}, \bibinfo {author} {\bibfnamefont {H.}~\bibnamefont {Nakamoto}},
		\bibinfo {author} {\bibfnamefont {Y.}~\bibnamefont {Nose}}, \bibinfo {author}
		{\bibfnamefont {T.}~\bibnamefont {Uda}}, \ and\ \bibinfo {author}
		{\bibfnamefont {Y.}~\bibnamefont {Shirai}},\ }\href {\doibase
		10.1002/pssc.201400291} {\bibfield  {journal} {\bibinfo  {journal} {Phys.
				Status Solidi C}\ }\textbf {\bibinfo {volume} {12}},\ \bibinfo {pages} {520}
		(\bibinfo {year} {2015})}\BibitemShut {NoStop}%
	\bibitem [{\citenamefont {Young}\ \emph {et~al.}(2000)\citenamefont {Young},
		\citenamefont {Coutts}, \citenamefont {Kaydanov}, \citenamefont {Gilmore},\
		and\ \citenamefont {Mulligan}}]{2000-JVTASF-band}%
	\BibitemOpen
	\bibfield  {author} {\bibinfo {author} {\bibfnamefont {D.~L.}\ \bibnamefont
			{Young}}, \bibinfo {author} {\bibfnamefont {T.~J.}\ \bibnamefont {Coutts}},
		\bibinfo {author} {\bibfnamefont {V.~I.}\ \bibnamefont {Kaydanov}}, \bibinfo
		{author} {\bibfnamefont {A.~S.}\ \bibnamefont {Gilmore}}, \ and\ \bibinfo
		{author} {\bibfnamefont {W.~P.}\ \bibnamefont {Mulligan}},\ }\href {\doibase
		10.1116/1.1290372} {\bibfield  {journal} {\bibinfo  {journal} {J. Vac. Sci.
				Technol. A}\ }\textbf {\bibinfo {volume} {18}},\ \bibinfo {pages} {2978}
		(\bibinfo {year} {2000})}\BibitemShut {NoStop}%
	\bibitem [{\citenamefont {Zhang}\ \emph {et~al.}(2009)\citenamefont {Zhang},
		\citenamefont {Mascarenhas}, \citenamefont {Wei},\ and\ \citenamefont
		{Wang}}]{2009-PRB-LRO}%
	\BibitemOpen
	\bibfield  {author} {\bibinfo {author} {\bibfnamefont {Y.}~\bibnamefont
			{Zhang}}, \bibinfo {author} {\bibfnamefont {A.}~\bibnamefont {Mascarenhas}},
		\bibinfo {author} {\bibfnamefont {S.-H.}\ \bibnamefont {Wei}}, \ and\
		\bibinfo {author} {\bibfnamefont {L.-W.}\ \bibnamefont {Wang}},\ }\href
	{\doibase 10.1103/PhysRevB.80.045206} {\bibfield  {journal} {\bibinfo
			{journal} {Phys. Rev. B}\ }\textbf {\bibinfo {volume} {80}},\ \bibinfo
		{pages} {045206} (\bibinfo {year} {2009})}\BibitemShut {NoStop}%
	\bibitem [{\citenamefont {Hohenberg}\ and\ \citenamefont {Kohn}(1964)}]{DFT-1}%
	\BibitemOpen
	\bibfield  {author} {\bibinfo {author} {\bibfnamefont {P.}~\bibnamefont
			{Hohenberg}}\ and\ \bibinfo {author} {\bibfnamefont {W.}~\bibnamefont
			{Kohn}},\ }\href {\doibase 10.1103/PhysRev.136.B864} {\bibfield  {journal}
		{\bibinfo  {journal} {Phys. Rev.}\ }\textbf {\bibinfo {volume} {136}},\
		\bibinfo {pages} {B864} (\bibinfo {year} {1964})}\BibitemShut {NoStop}%
	\bibitem [{\citenamefont {Kohn}\ and\ \citenamefont {Sham}(1965)}]{DFT-2}%
	\BibitemOpen
	\bibfield  {author} {\bibinfo {author} {\bibfnamefont {W.}~\bibnamefont
			{Kohn}}\ and\ \bibinfo {author} {\bibfnamefont {L.~J.}\ \bibnamefont
			{Sham}},\ }\href {\doibase 10.1103/PhysRev.140.A1133} {\bibfield  {journal}
		{\bibinfo  {journal} {Phys. Rev.}\ }\textbf {\bibinfo {volume} {140}},\
		\bibinfo {pages} {A1133} (\bibinfo {year} {1965})}\BibitemShut {NoStop}%
	\bibitem [{\citenamefont {Kresse}\ and\ \citenamefont {Hafner}(1993)}]{VASP-1}%
	\BibitemOpen
	\bibfield  {author} {\bibinfo {author} {\bibfnamefont {G.}~\bibnamefont
			{Kresse}}\ and\ \bibinfo {author} {\bibfnamefont {J.}~\bibnamefont
			{Hafner}},\ }\href {\doibase 10.1103/PhysRevB.47.558} {\bibfield  {journal}
		{\bibinfo  {journal} {Phys. Rev. B}\ }\textbf {\bibinfo {volume} {47}},\
		\bibinfo {pages} {558} (\bibinfo {year} {1993})}\BibitemShut {NoStop}%
	\bibitem [{\citenamefont {Kresse}\ and\ \citenamefont {Hafner}(1994)}]{VASP-2}%
	\BibitemOpen
	\bibfield  {author} {\bibinfo {author} {\bibfnamefont {G.}~\bibnamefont
			{Kresse}}\ and\ \bibinfo {author} {\bibfnamefont {J.}~\bibnamefont
			{Hafner}},\ }\href {\doibase 10.1103/PhysRevB.49.14251} {\bibfield  {journal}
		{\bibinfo  {journal} {Phys. Rev. B}\ }\textbf {\bibinfo {volume} {49}},\
		\bibinfo {pages} {14251} (\bibinfo {year} {1994})}\BibitemShut {NoStop}%
	\bibitem [{\citenamefont {Kresse}\ and\ \citenamefont
		{Furthm\"uller}(1996)}]{VASP-3}%
	\BibitemOpen
	\bibfield  {author} {\bibinfo {author} {\bibfnamefont {G.}~\bibnamefont
			{Kresse}}\ and\ \bibinfo {author} {\bibfnamefont {J.}~\bibnamefont
			{Furthm\"uller}},\ }\href {\doibase 10.1103/PhysRevB.54.11169} {\bibfield
		{journal} {\bibinfo  {journal} {Phys. Rev. B}\ }\textbf {\bibinfo {volume}
			{54}},\ \bibinfo {pages} {11169} (\bibinfo {year} {1996})}\BibitemShut
	{NoStop}%
	\bibitem [{\citenamefont {Sun}\ \emph {et~al.}(2015)\citenamefont {Sun},
		\citenamefont {Ruzsinszky},\ and\ \citenamefont {Perdew}}]{2015-PRL-SCAN}%
	\BibitemOpen
	\bibfield  {author} {\bibinfo {author} {\bibfnamefont {J.}~\bibnamefont
			{Sun}}, \bibinfo {author} {\bibfnamefont {A.}~\bibnamefont {Ruzsinszky}}, \
		and\ \bibinfo {author} {\bibfnamefont {J.~P.}\ \bibnamefont {Perdew}},\
	}\href {\doibase 10.1103/PhysRevLett.115.036402} {\bibfield  {journal}
		{\bibinfo  {journal} {Phys. Rev. Lett.}\ }\textbf {\bibinfo {volume} {115}},\
		\bibinfo {pages} {036402} (\bibinfo {year} {2015})}\BibitemShut {NoStop}%
	\bibitem [{\citenamefont {Isaacs}\ and\ \citenamefont
		{Wolverton}(2018)}]{2018-PRM-SCAN}%
	\BibitemOpen
	\bibfield  {author} {\bibinfo {author} {\bibfnamefont {E.~B.}\ \bibnamefont
			{Isaacs}}\ and\ \bibinfo {author} {\bibfnamefont {C.}~\bibnamefont
			{Wolverton}},\ }\href {\doibase 10.1103/PhysRevMaterials.2.063801} {\bibfield
		{journal} {\bibinfo  {journal} {Phys. Rev. Mater.}\ }\textbf {\bibinfo
			{volume} {2}},\ \bibinfo {pages} {063801} (\bibinfo {year}
		{2018})}\BibitemShut {NoStop}%
	\bibitem [{\citenamefont {Heyd}\ \emph {et~al.}(2003)\citenamefont {Heyd},
		\citenamefont {Scuseria},\ and\ \citenamefont {Ernzerhof}}]{HSE06}%
	\BibitemOpen
	\bibfield  {author} {\bibinfo {author} {\bibfnamefont {J.}~\bibnamefont
			{Heyd}}, \bibinfo {author} {\bibfnamefont {G.~E.}\ \bibnamefont {Scuseria}},
		\ and\ \bibinfo {author} {\bibfnamefont {M.}~\bibnamefont {Ernzerhof}},\
	}\href {\doibase 10.1063/1.1564060} {\bibfield  {journal} {\bibinfo
			{journal} {J. Chem. Phys.}\ }\textbf {\bibinfo {volume} {118}},\ \bibinfo
		{pages} {8207} (\bibinfo {year} {2003})}\BibitemShut {NoStop}%
	\bibitem [{\citenamefont {Seryogin}\ \emph {et~al.}(1999)\citenamefont
		{Seryogin}, \citenamefont {Nikishin}, \citenamefont {Temkin}, \citenamefont
		{Mintairov}, \citenamefont {Merz},\ and\ \citenamefont
		{Holtz}}]{1999-APL-ZnSnP2-expt}%
	\BibitemOpen
	\bibfield  {author} {\bibinfo {author} {\bibfnamefont {G.~A.}\ \bibnamefont
			{Seryogin}}, \bibinfo {author} {\bibfnamefont {S.~A.}\ \bibnamefont
			{Nikishin}}, \bibinfo {author} {\bibfnamefont {H.}~\bibnamefont {Temkin}},
		\bibinfo {author} {\bibfnamefont {A.~M.}\ \bibnamefont {Mintairov}}, \bibinfo
		{author} {\bibfnamefont {J.~L.}\ \bibnamefont {Merz}}, \ and\ \bibinfo
		{author} {\bibfnamefont {M.}~\bibnamefont {Holtz}},\ }\href {\doibase
		10.1063/1.123778} {\bibfield  {journal} {\bibinfo  {journal} {Appl. Phys.
				Lett.}\ }\textbf {\bibinfo {volume} {74}},\ \bibinfo {pages} {2128} (\bibinfo
		{year} {1999})}\BibitemShut {NoStop}%
	\bibitem [{Git()}]{Github-code}%
	\BibitemOpen
	\href
	{https://github.com/HanpuLiang/special-quasidisorder-structure-code-SQDS}
	{}\bibinfo {note}
	{Https://github.com/HanpuLiang/special-quasidisorder-structure-code-SQDS}\BibitemShut
	{NoStop}%
\end{thebibliography}
%

\end{document}